\documentclass{article}

\usepackage{longtable}
\usepackage{supertabular}
\usepackage{multirow} 
\usepackage{floatrow}
\usepackage{array}
\usepackage{soul}
\usepackage{float}
\usepackage{graphicx}
\usepackage{caption}
\usepackage{subcaption}
\usepackage[table]{xcolor}
\usepackage{pbox}
\newcolumntype{P}[1]{>{\centering\arraybackslash}p{#1}}

    \PassOptionsToPackage{numbers, compress}{natbib}

    \usepackage[preprint]{neurips_2020}

\usepackage[utf8]{inputenc} 
\usepackage[T1]{fontenc}    
\usepackage{hyperref}       
\usepackage{url}            
\usepackage{booktabs}       
\usepackage{amsfonts}       
\usepackage{nicefrac}       
\usepackage{microtype}      
\usepackage{graphicx}
\usepackage{subcaption}
\usepackage{rotating}
\usepackage{titlesec}
\usepackage[inline]{enumitem}

\newlist{paragraphs}{itemize*}{1}
\setlist[paragraphs]{
   label=(\textbf{\thesection}),
   itemjoin=\newline\hspace*{\parindent}
}
\makeatletter
\def\namedlabel#1#2#3{\begingroup
    #3%
    \def\@currentlabel{#2}%
    \phantomsection\label{#1}\endgroup
}
\makeatother
\usepackage{dsfont}

\usepackage{amsmath,amsfonts,bm}

\def\eqref#1{equation~\ref{#1}}

\def\1{\bm{1}}

\DeclareMathAlphabet{\mathsfit}{\encodingdefault}{\sfdefault}{m}{sl}
\SetMathAlphabet{\mathsfit}{bold}{\encodingdefault}{\sfdefault}{bx}{n}

\usepackage{semicrunch}
\usepackage{wrapfig,lipsum,booktabs}
\usepackage{hyperref}
\usepackage{xspace}
\usepackage[utf8]{inputenc}
\usepackage[english]{babel}
\usepackage{amsthm}
\usepackage{amssymb}
\usepackage{subcaption}
\usepackage{floatrow}
\usepackage{mathtools}
\usepackage{hyperref}
\usepackage{xcolor}
\usepackage{appendix}
\usepackage[normalem]{ulem}

\usepackage{etoolbox}
\makeatletter
\patchcmd{\hyper@makecurrent}{%
    \ifx\Hy@param\Hy@chapterstring
        \let\Hy@param\Hy@chapapp
    \fi
}{%
    \iftoggle{inappendix}{
        \@checkappendixparam{chapter}%
        \@checkappendixparam{section}%
        \@checkappendixparam{subsection}%
        \@checkappendixparam{subsubsection}%
        \@checkappendixparam{paragraph}%
        \@checkappendixparam{subparagraph}%
    }{}%
}{}{\errmessage{failed to patch}}

\newcommand*{\@checkappendixparam}[1]{%
    \def\@checkappendixparamtmp{#1}%
    \ifx\Hy@param\@checkappendixparamtmp
        \let\Hy@param\Hy@appendixstring
    \fi
}
\makeatletter

\newtoggle{inappendix}
\togglefalse{inappendix}

\apptocmd{\appendix}{\toggletrue{inappendix}}{}{\errmessage{failed to patch}}
\apptocmd{\subappendices}{\toggletrue{inappendix}}{}{\errmessage{failed to patch}}

\newcommand\strike{\bgroup\markoverwith{\textcolor{red}{\rule[0.5ex]{2pt}{0.4pt}}}\ULon}
\definecolor{darkgreen}{RGB}{0,180,56}

\newtheorem*{rep@theorem}{\rep@title}
\newcommand{\newreptheorem}[2]{%
\newenvironment{rep#1}[1]{%
 \def\rep@title{#2 \ref{##1}}%
 \begin{rep@theorem}}%
 {\end{rep@theorem}}}
\makeatother
\newreptheorem{theorem}{Theorem}
\newreptheorem{lemma}{Lemma}
\newreptheorem{assumption}{Assumption}

\usepackage{graphicx}
\usepackage{wrapfig}
\usepackage{algorithm}
\usepackage{algcompatible}
\usepackage{multicol}
\usepackage{enumitem}

\title{Lattice Thermal Conductivity Prediction using Symbolic Regression and Machine Learning}

\author{%
  Christian Loftis\\
  Department of Computer Science and Engineering\\
  University of South Carolina, Columbia, SC, 29201, USA\\
   \And
   Kunpeng Yuan\\
  Department of Mechanical Engineering\\
  University of South Carolina, Columbia, SC, 29201, USA\\
  Key Laboratory of Ocean Energy Utilization and Energy Conservation of Ministry of Education\\
School of Energy and Power Engineering,Dalian University of Technology\\
Dalian, 116024, China\\
  \And
  Yong Zhao\\
  Department of Computer Science and Engineering\\
  University of South Carolina, Columbia, SC, 29201, USA\\
   \And
   Ming Hu*\\
  Department of Mechanical Engineering\\
  University of South Carolina, Columbia, SC, 29201, USA\\
  \And
  Jianjun Hu *\\
  Department of Computer Science and Engineering\\
  University of South Carolina\\
  \texttt{jianjunh@cse.sc.edu}\\
}

\begin{document}

\maketitle

\begin{abstract}
Prediction models of lattice thermal conductivity ($\kappa_{L}$) have wide applications in the discovery of thermoelectrics, thermal barrier coatings, and thermal management of semiconductors. $\kappa_{L}$ is notoriously difficult to predict. While classic models such as the Debye-Callaway model and the Slack model have been used to approximate the $\kappa_{L}$ of inorganic compounds, their accuracy is far from being satisfactory. Herein, we propose a genetic programming based Symbolic Regression (SR) approach for explicit $\kappa_{L}$ models and compare it with Multi-Layer Perceptron neural networks and a Random Forest Regressor using a hybrid cross-validation approach including both K-Fold CV and holdout validation. Four formulae have been discovered by our symbolic regression approach that outperform the Slack formula as evaluated on our dataset. Through the analysis of our models’ performance and the formulae generated, we found that the trained formulae successfully reproduce the correct physical law that governs the lattice thermal conductivity of materials. We also identified that extrapolation prediction remains to be a key issue in both symbolic regression and regular machine learning methods and find the distribution of the samples place a key role in training a prediction model with high generalization capability.
\end{abstract}

\section{Introduction}

Having the capability to predict lattice thermal conductivity ($\kappa_{L}$) of a crystalline material based on its composition and structure information has wide applications in new materials discovery, and thus has received noticeable attention in the thermodynamics field \cite{slack-morelli-paper,callaway_citation,georgiatechpaper}. It enables materials scientists to screen materials with desired $\kappa_{L}$ without having to synthesize the materials first for testing. The advantages of materials with both high and low $\kappa_{L}$ abound. For example, materials with high $\kappa_{L}$ are desirable for conducting heat, and their uses range from being used for coolant pipes in nuclear power plants to being used for heat sinks. $\kappa_{L}$ is especially important for semiconductors, whose electrical resistance rise as their temperature falls. Being able to optimize for thermal conductivity independently of electron conductivity enables materials researchers to create electrical insulators that conduct heat well, or inversely, to create electrical conductors that do not transfer heat well. Possessing this degree of control over a material's conductivity (both thermal and electric) will allow researchers to synthesize materials for use in electronics that can transmit electricity easily yet conduct less heat than other materials with the same electrical conductivity. This leads to electronics that do not overheat as quickly, despite high transistor density. Slack and Morelli state that “Its manipulation and control have impacted an enormous variety of technical applications, including thermal management of mechanical, electrical, chemical, and nuclear systems; thermal barriers and thermal insulation materials; more efficient thermoelectric materials; and sensors and transducers." \cite{slack-morelli-paper}

The thermodynamics research field has contributed several analytical models for calculating lattice thermal conductivity of materials including the well-known Slack model (Formula \ref{eq:slack-berman}) \cite{slack-morelli-paper} and Debye-Callaway Model (Formula \ref{eq:debye-callaway}) \cite{callaway_citation}.
\\

\begin{equation}
    \kappa_{L} = A \cdot \frac{(\theta e)^3 M \sqrt[3]{Vn_p}}{n^{4/3} T \gamma_{a}^2}\label{eq:slack-berman}
\end{equation}
\\
\begin{equation}
    \kappa_{L, tot} = A_{1}\frac{Mv_{s}^{3}}{TV^{2/3}n^{1/3}} + A_{2}\frac{v_{s}}{V^{2/3}}(1-\frac{1}{n^{2/3}})\label{eq:debye-callaway}
\end{equation}

While these models are insightful, comparison with experimentally measured thermal conductivity has indicated that there are still plenty room for improvement \cite{ma2014examining,allen2013improved,nath2017high}. They have also been shown to be less accurate than machine learning models that have been developed to predict $\kappa_{L}$ \cite{georgiatechpaper,callaway_citation}. 

Recently, several studies have applied machine learning (ML) methods for thermal conductivity prediction \cite{georgiatechpaper,tawfik2020predicting,juneja2020guided,zhu2020charting, wang2020identification, wan2019materials,wei2020genetic,yan2020seeking,yamada2019predicting}. Chen et. al. \cite{georgiatechpaper} propose a Gaussian Process Regression combined with feature engineering by recursive feature elimination and Random Forest based feature selection for LTC prediction. When applied to the small data set of 100 samples, they report a performance of $R_2$~0.93 when trained on 76 samples and tested on 19 samples. However, this result is questionable and may be due to the high redundancy/similarity of the samples. To improve the generalization performance, Juneja and Singh \cite{juneja2020guided} propose a localized regression based patchwork kriging approach with elemental and structural descriptors for $\kappa_{L}$ prediction. When applied to a dataset of 2838 materials, higher transferability has been achieved. Wan et al. \cite{wan2019materials} applied XGBoost algorithm based on the descriptors of crystal structural and compositional information to $\kappa_{L}$ prediction. Two geometric descriptors have also been shown to be closely related to thermal conductivities \cite{wei2020genetic}. To address the issue of limited materials with annotated $\kappa_{L}$, a shortgun transfer learning approaches has been proposed and applied to a small $\kappa_{L}$ dataset of 95 samples. A major improvement of $\kappa_{L}$ prediction comes from Zhu et al.'s work \cite{zhu2020charting}, in which both Graph convolution network and Random Forest with elemental and structure features have been used to derive a prediction model over a much larger dataset with ~2700 training samples. However, all these studies have not evaluated the real extrapolation performance \cite{xiong2020evaluating}. While these machine learning approaches have demonstrated themselves to be suited to predicting $\kappa_{L}$, but they are unfortunately limited by nature in the insight that they can provide to the thermal science community as most of the ML models are essentially based on interpolation. 


This study seeks to bridge the gap between the analytical models and machine learning models for $\kappa_{L}$ prediction by exploring three types of models by focusing on the extrapolative prediction or generalization performance of three types of prediction models. Our first model is based on genetic programming Symbolic Regression (SR), which is an evolutionary algorithm that can generate formulae to map ordinal material properties to $\kappa_{L}$. The second model is a deep neural network model using a Multi-Layer Perceptron (MLP) powered by the Adam optimizer to predict $\kappa_{L}$ by analyzing both the linear and nonlinear relationships in the data. Finally, the third model uses the Random Forest Regressor (RFR), a traditional machine learning method that has been shown to be effective in predicting $\kappa_{L}$ \cite{low-thermal-conductivity, zhu2020charting,georgiatechpaper}. We derive several formulae using the symbolic regression method that outperform the Slack formula on our test dataset. In addition, analysis of our models’ performance and formulae highlight interesting variable relationships to $\kappa_{L}$ calculation and prediction, which showed the advantage of interpretable models of symbolic regression.

The Symbolic Regression models in this study take three forms. The first form, referred to as GP1, uses a limited function set with the intention of discovering models similar to the classic Slack model. The second, GP2, is provided with a richer function set to find formulae that are better than the Slack formula or are otherwise analytically distinct. Finally, the third model is a proof-of-concept model that illustrates the effectiveness of the Symbolic Regression methodology by attempting to rediscover the Slack formula from raw data points.

\section{Methods}

\subsection{Dataset and features}

\begin{figure}[H]
\centerline{\includegraphics[width=5in, height=4.131in]{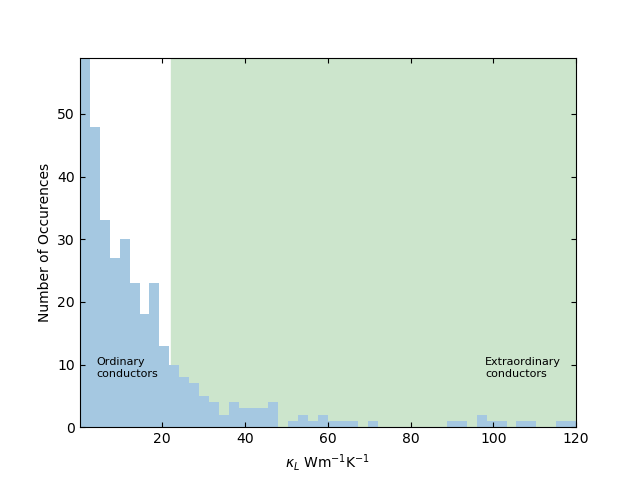}}
\vspace*{8pt}
\caption{A histogram depicting the range and distribution of the $\kappa_{L}$ of the materials in the dataset, with each bar representing a precision of 2.4 $Wm^{-1}K^{-1}$. The X axis displays the $\kappa_{L}$ of the materials, while the Y axis shows how many samples in the dataset fall within this range. The top 20\% of the $\kappa_{L}$ range (highlighted here in green) is excluded from the training set in select experiments.}
\label{fig:distribution}
\end{figure}

Each model is provided with the same set of descriptors (Table \ref{table:descriptors}), with the exception that the symbolic regression models are unable to use the space group variable. This is due to the nature of the SR models, which require numeric fields that can be used as variables inside of formulae. In order to mitigate issues with fitting the models to the data, all materials with observed $\kappa_{L}$ above 120 are recognized as outliers and thus trimmed from the dataset used for training and validation. The value 120 is chosen because the dataset contained a much higher concentration of data points with $\kappa_{L}$ immediately below 120 than those above 120. The distribution and range of the dataset's $\kappa_{L}$ values can be seen in Figure \ref{fig:distribution}. In total, there are 347 samples.

To prepare the dataset, all the first-principles calculations are carried out based on density functional theory (DFT) as implemented in the Vienna ab-initio Simulation Package (VASP) \cite{PhysRevB.54.11169}. The projector-augmented wave (PAW) pseudopotentials \cite{PhysRevB.59.1758} are used to describe the interaction among atoms and the generalized gradient approximation (GGA) in the Perdew-Burke-Ernzerhof (PBE) \cite{PhysRevLett.77.3865} form is chosen as the exchange-correlation functional. The kinetic energy cutoff of the plane-wave function is set as the default maximum energy cutoff for each material. A Monkhorst-Pack \cite{PhysRevB.13.5188} k-point grid of 0.4 2$\pi$/Å is used to sample the first Brillouin zone. The convergent criterion for the total energy difference between two successive self-consistency steps is $10^{-5}$ eV and all the geometries are fully relaxed until the maximum force acting on each atom is less than 0.01 eV/Å. The elastic constants are calculated from the strain-stress relationship. According to the Voigt-Reuss-Hill theory \cite{Toonder_1999}, the corresponding elastic properties, such as the bulk modulus B and shear modulus G, can be evaluated from the elastic constants. To obtain the Grüneisen parameter, we calculate the change in the elastic properties with volume by changing the volume from -1.5 \% to 1.5\% (5 points in total) \cite{PhysRevB.95.155206}.

\begin{table}[th]
\centering
{\begin{tabular}{@{}|c|c|@{}} \toprule
\hline
Variable Symbol & Definition\\
\hline
$V$ & Volume per atom \\ \hline
$T$ & Temperature (constant: 300 $K$) \\ \hline
$M$ & Average atomic mass \\ \hline
$n$ & Total number of atoms in unit cell \\ \hline
$n_{p}$ & Total number of atoms in primitive  cell \\ \hline
$B$ & Bulk modulus calculated from $C_{ij}$\footnote{Second-order elastic constants} \\ \hline
$G$ & Shear modulus calculated from $C_{ij}$ \\ \hline
$E$ & Young’s modulus \\ \hline
$\nu$ & Poisson’s ratio \\ \hline
$H$ & Estimated hardness \\ \hline
$B'$ & ($\delta$ B/$\delta.$V) \\ \hline
$G'$ & ($\delta$ G/$\delta.$V) \\ \hline
$\rho$ & Mass density \\ \hline
$v_{L}$ & Sound velocities of the longitude \\ \hline
$v_{S}$ & Sound velocities of the shear \\ \hline
$v_{a}$ & Corresponding average velocity \\ \hline
$\Theta_{D}$ & Debye temperature \\ \hline
$\gamma_{L}$ & Longitude acoustic Grüneisen parameters \\ \hline
$\gamma_{S}$ & Shear acoustic Grüneisen parameters \\ \hline
$\gamma_{a}$ & Average acoustic Grüneisen parameters \\ \hline
$A$ & Empirical parameter \\  \hline
\end{tabular}}
\caption{\label{tab:list_of_variables} List of descriptors and their respective definitions.}
\label{table:descriptors}
\end{table}

\subsection{Preprocessing}

The space group descriptor is converted into binary encoding in order for the MLP model to be able to process the categorical value properly. In addition to this, the fields shown in Table \ref{tab:list_of_variables} are scaled using min-max scaling to restrict all variables to a minimum and maximum of 0 and 1. The RF regression model requires that the space group descriptor be converted to ordinal values that represent the various space groups. The dataset needs no preprocessing modification for the SR model, aside from the removal of the space group descriptor.

The architectures for the various models are described below. Barring the Symbolic Regression models, there is only one architecture used to create each model.
\subsection{Symbolic Regression}

Symbolic Regression is a form of regression that uses mathematical operators as building blocks to intelligently create formulae, with two objectives: minimizing the prediction error, and maximizing the simplicity of the formulae produced. To accomplish this, it uses the concept of the Pareto Frontier\cite{pareto-frontier-article} to optimize both attributes simultaneously. Through producing simple formulae, Symbolic Regression substantially decreases the likelihood of overfitting to latent trends in the dataset that do not generalize. This is particularly applicable to the fields of physics and material science, as most of the physical laws, when expressed as equations, are relatively mathematically simple. Examples include $F=ma$ and $E=mc^2$. 

\begin{figure}[H]
\centering
\begin{tabular}{P{40mm}P{40mm}P{40mm}}
  \includegraphics[width=40mm]{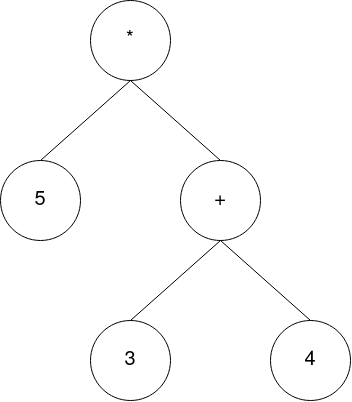} & \includegraphics[width=40mm]{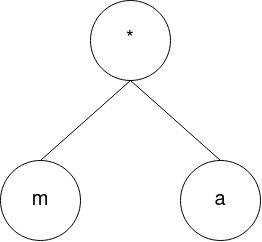} & \includegraphics[width=40mm]{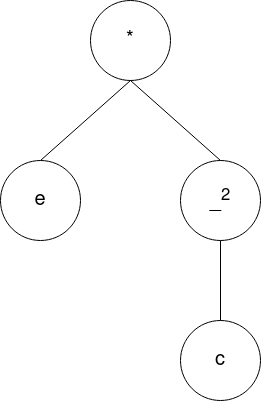} \\
(a) $5 * (3 + 4)$ & (b) $F=ma$ & (c) $E=mc^{2}$ \\[6pt]
\end{tabular}
\caption{\label{fig:sample_trees}Examples of function trees to be represented in the symbolic regression algorithm. The algorithm creates trees similar to the ones above, using the descriptors in the dataset as variables.}
\end{figure}

Our methodology for creating these formulae is through the genetic programming (GP) approach\cite{banzhaf1998genetic}. In GP, formulae are represented as unique function trees (see Figure \ref{fig:sample_trees} for example), with operators, input variables/descriptors, and constants as nodes in each individual tree. They are evaluated for their performance on the dataset and their simplicity, and using the Pareto frontier a fitness value is generated for each model. The models are then compared to each other by fitness rankings, with the fitter models having higher probability to proceed onto the next stage of the evolution process: crossover and mutation. Subtrees are randomly selected from two partner trees, and offspring in the form of permutations of the parents are created through this crossover stage. In this way, we use natural selection to select successful traits from parents, and pass them down to offspring. From here, the genetic process is repeated for a set number of generations, and the most successful formula is returned. Genetic programming provides a method for allowing beneficial traits and terms to remain in the function while simultaneously discarding unhelpful terms from the equation. It does not guarantee a perfect solution, but rather through exploring several partial solutions, it is able to intelligently combine them together to create a unified formula for approximating a function that lies underneath a dataset.

Symbolic regression has a few disadvantages when compared to other ML approaches, but it has one unique advantage that standard ML is unable to replicate. Symbolic regression generates a human-readable function in a mathematical notation. This formula can be analyzed to derive physical insight into the processes that drive the subject to behave the way that it does. As aforementioned, symbolic regression is not a flawless method. It has several disadvantages, among which are computational inefficiency during training and size of the search space. GP based Symbolic Regression is notoriously computationally inefficient, drawing much more resources for the training phase than statistics-based ML models require. However, once the formulae are produced, they can be run instantly based on their respective complexity, which the algorithm aims to reduce throughout its process. 

The size of the search space is of much more concern when applying Symbolic Regression versus most statistics-based ML algorithms. The search space for a Symbolic Regression algorithm is theoretically infinite, as there are infinite formulae that can be produced from the genetic programming functions (GP functions) provided to the algorithm. The odds of the algorithm finding and settling for the formula that perfectly maps the provided fields to the desired output is low. In order to offset this large search space, we restrict the height that the function trees are permitted to obtain. This places a finite capacity on the amount of formulae that can be generated while simultaneously ensuring that the formulae we generate remain below a maximum complexity threshold. As aforementioned, physical formulae are mathematically simple, so it is a safe way to prune the search space. To narrow down the search space even farther, we restrict the function set that the algorithm is allowed to make use of. The function sets for these two models are described in Table \ref{tab:sr_model_architectures} below. The number of functions provided to the model has a direct correlation to the size of the search space; therefore, by limiting the GP function set, the dimensionality of the problem is reduced, and the likelihood of convergence is increased.

This experiment explores two methodologies for calculating $\kappa_{L}$ through Symbolic Regression, as described in Table \ref{tab:sr_model_architectures}. In addition, it uses a third methodology to prove the validity of the Symbolic Regression algorithm for this dataset. The implementation for the SR Model was provided by the FastSR library\cite{cfusting}.

\subsection{Verifying Effectiveness of Symbolic Regression}

In order to provide a benchmark for the validity of our Symbolic Regression algorithm and demonstrate its ability to learn from a dataset, we created a separate experiment in which a Symbolic Regression model is allowed to train from the Slack predictions for the dataset provided. We provided the Symbolic Regression model with the V, M, $\theta_{D}$, $\gamma_{a}$, n, $n_p$, A, and T variables, and let it view the Slack model calculated $\kappa_{L}$ in order to learn. The goal of this experiment was to demonstrate the learning capacity of our Symbolic Regression methodology by allowing it to train on the Slack predictions, and see how closely it can approximate the Slack formula through exposure to the variables that the Slack formula uses.

The model was permitted to use the following GP functions: 

$\times, \div, f^{-1}(x), ln(|x|), e^{x}, \sqrt[3](x), x^{3}, \sqrt[2](x), x^{2},$ 

\begin{figure}[H]
\centerline{\includegraphics[width=5in, height=4.131in]{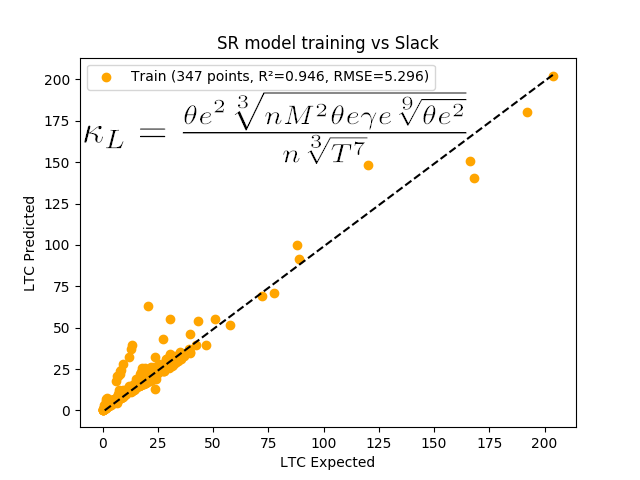}}
\vspace*{8pt}
\caption{Parity plot for the evolved formula \ref{eq:slack-reproduction}}
\label{fig:sr_reproduction_parity_plot}
\end{figure}

In addition, randomly generated constants following a Gaussian distribution with $\mu$=0 and $\sigma$=10 are also provided to evolve the coefficients in the formulae. The algorithm created 1,000 generations of 1,500 formulae. The model is limited to producing formula trees with a maximum height of 7, and 5 of the 1500 functions introduced with each generation were generated completely randomly, in order to prevent the model from fixating on a local minimum in the cost function gradient. Ultimately, our SR algorithm evolved a formula (see Formula \ref{eq:slack-reproduction}) that is extremely close to that of the Slack formula, its target. The evolved formula achieves a RMSE of 5.296 and $R_2$ of 0.946. The parity plot of the predicted $\kappa_{L}$ versus Slack model values are shown in Figure \ref{fig:sr_reproduction_parity_plot}. This results reflect the ability of the evolved SR formula to map inputs to their predicted $\kappa_{L}$ Slack values.

\begin{equation}
    \kappa_{L} = \frac{\theta e^{2}\sqrt[3]{n M^{2} \theta e \gamma e \sqrt[9]{\theta e^{2}}}}{n\sqrt[3]{T^{7}}} \label{eq:slack-reproduction}
\end{equation}

Formula \ref{eq:slack-reproduction} is the simplified form of the evolved formula, which boasts a very high $R^{2}$ score of 0.946, meaning that it very closely mirrors the Slack equation. Interestingly, Formula \ref{eq:slack-reproduction} does not use the A, V, and $n_p$ variables. Despite this, it still obtains an effective approximation. Another interesting observation is that Formula \ref{eq:slack-reproduction} places the $\gamma_{a}$ variable in the numerator rather than the denominator, and changes the exponent from 2 to $\frac{1}{3}$. This creates a relationship where the $\kappa_{L}$ and the $\gamma_{a}$ values have a partial direct correlation. The $\theta_{D}$ variable in the numerator of Formula \ref{eq:slack-berman} has an exponent of 3, yet the same variable in Formula \ref{eq:slack-reproduction}, through some algebraic manipulation, has an exponent of $2 + \frac{2}{27}$, meaning that Formula \ref{eq:slack-reproduction} places less importance on the Debye temperature than the original Slack equation. These changes in scaling could be a result of the model compensating for the missing variables, thus demonstrating the plastic and adaptive nature of the Symbolic Regression algorithm. The fact that it is able to reproduce the Slack equation with an $R^{2}$ value of 0.946 means that it has the potential to regress a formula with a comparable coefficient of determination with the actual $\kappa_{L}$ values set as the supervised learning set.

\begin{table}[H]
{\begin{tabular}{|p{7cm}|p{9cm}|} \toprule
\hline
GP1 & GP2\\
\hline
$\times$, $\div$, $f^{-1}(x)$, random constants on a Gaussian distribution with $mu$=0 and $\sigma$=10 & \pbox{20cm}{$\times$, $\div, f^{-1}(x), ln(|x|), e^{x}, x^{2}, x^{3}, \sqrt[2](x), \sqrt[3](x), sin(x)$,\\ $cos(x), tan(x)$, random constants on a Gaussian\\ distribution with $\mu$=0 and $\sigma$=10} \\
\hline
500 Generations & 500 Generations \\
\hline
2000 Population size & 2000 Population size \\
\hline
7 Max height & 10 Max height \\
\hline
30\% Mutation probability & 30\% Mutation probability \\
\hline
70\% Crossover probability & 70\% Crossover probability \\
\hline
\end{tabular}}
\caption{\label{tab:sr_model_architectures} Configurations of two SR Model architectures.}
\end{table}

\subsection{Multi-Layer Perceptron (MLP) Neural Network}

Neural networks are mathematical models that take in a predefined amount of inputs and convert them through multiple layers or linear or nonlinear transformation to generate a predefined number of outputs. Through stacking the artificial neurons in layers that feed their outputs forward through the network, and using optimization algorithms such as stochastic gradient descent or the Adam optimizer, the weights on these neurons are adjusted to be able to output values close to those in the training set. It is well known that deep neural networks are excellent at learning nonlinear relationships \cite{goodfellow2016deep}, but deep learning approaches such as the MLP require vast amounts of data to effectively learn trends and relationships. In addition, MLP models form a black box system that, while accurate, is unable to provide scientists with insight into how the model is able to map the input variables to their expected output variables; they are a tool that can be used but their processes for reaching their solutions cannot be understood easily despite recent efforts for explainable deep neural network models\cite{hu2018explainable}.

Many optimization algorithms require the data to be passed over multiple times in order for trends to be accurately learned, with one forward and backward pass constituting what is referred to as an epoch. The amount of epochs needed to accurately learn trends is inversely related to the size of the training set and directly related to the complexity of the problem. In order to ensure that our model is able to adequately learn from the dataset, we allowed the model to train over 30 epochs for each step in the 5-fold cross-validation process. To offset any overfitting which may manifest as a result of this process, we make use of random dropout to address the issue, in which connections between neurons are randomly deactivated when the architecture is compiled.

\begin{figure}[H]
\centerline{\includegraphics[width=5in]{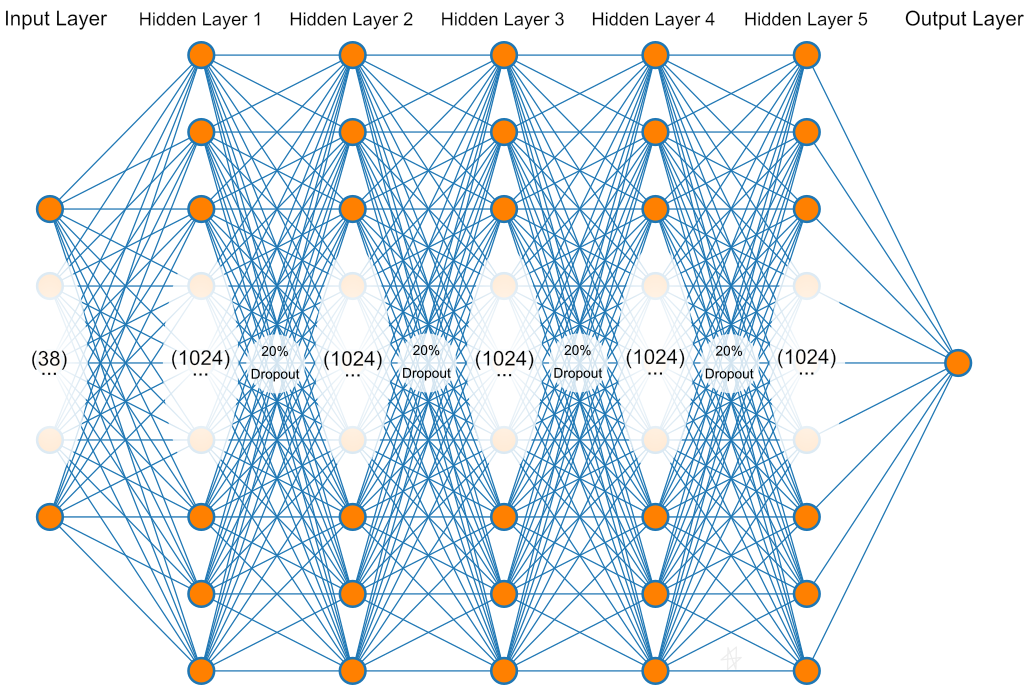}}
\vspace*{8pt}
\caption{Architecture of the MLP model}
\label{fig:nn_architecture}
\end{figure}

The MLP model, as depicted in Figure \ref{fig:nn_architecture}, makes use of 5 hidden layers with 1024 neurons in each layer, and a 20\% dropout between otherwise densely connected layers. ReLU (Rectified Linear Unit) is the activation function for all layers leading up to the final layer, which uses linear activation. The network was trained with MAE as the loss function, and makes use of the the Adam optimizer \cite{adam-paper}.

\subsection{Random Forest Regressor (RFR)}

Random Forest is an ensemble ML algorithm that takes advantage of a predefined number of decision trees. The dataset is divided up among the different decision trees, and each tree is given samples of the dataset through the bootstrap aggregation process. They are then trained on their individual samples, during which time the fitting process trims and prunes branches of the decision trees that are unnecessary or detrimental to the prediction of the final result. When the training process is complete and the model is provided with a query input data, each of the trees returns a prediction. In the case of a Random Forest Regressor, the model collects all of the predictions, calculates the average, and returns it as the final prediction.

Our RFR implementation uses a standard random forest model as provided by scikit-learn \cite{scikit-learn} Python package. The number of estimators is set as 100 and the loss function is set to MAE. By minimizing MAE rather than RMSE, the RFR model aims to provide a smooth prediction over all values, rather than overpunishing high residuals in $\kappa_{L}$ predictions.

\section{Experiments and Results}

\subsection{Training Process}
We conduct two types of experiments to compare the symbolic regression and ML models including the standard cross-validation tests and forward cross-validation extrapolation performance tests. 

For cross-validation experiments, during the training process, the data points are split into 10 equal subsets, and then 10-Fold cross-validation is performed. The data is split into sets of tenths, and one of these ten sets is hidden from the model and used as a validation set while the others are used for training. This process is repeated until each of the ten subsections are each used once as a validation set, and then the training phase is complete. The MLP was permitted to train for 30 epochs over the training set during each training interval of the cross validation process. 

For extrapolation test experiments, all models are trained on 80\% of the dataset, and then evaluated on a block of the remaining 20\% of the dataset in a process known as extrapolation testing \cite{xiong2020evaluating}. We also implement 5-fold cross validation on the training set, to reduce the chance of overfitting. For example, one of the extrapolation tests (depicted in Table \ref{tab:extrapolation-topbottom} and Figure \ref{fig:top_bottom_parity_plots}) sorts the materials in order of ascending $\kappa_{L}$, and allows the model to train on the middle 80\% of the data points. The bottom 10\% and the top 10\% are withheld from the model during the training phase and retained for the validation set. After the model has been trained, the model is tested on the validation set. If the model is able to perform similarly on both the training and validation sets, it is understood that the model has learned an underlying relationship between the input variables and $\kappa_{L}$. Because most of the test samples are not neighboring training samples in such tests, it is guaranteed that this relationship is not purely based on the sample's proximity to one another. The relationship learned can be used to predict $\kappa_{L}$ values of any sample, and thus must reflect the physical law that underlies $\kappa_{L}$ approximation.

\subsection{Random cross-validation Results}

\begin{table}[H]
\centering
{\begin{tabular}{|c|c|c|c|c|c|c|} \toprule
  & \textbf{GP1} & \textbf{GP2} & \textbf{MLP} & \textbf{RFR} & \textbf{Slack} & \textbf{Best}\\
\hline
\textbf{Formula} & $\frac{G}{H \cdot n_p}$ & $\frac{\sqrt[6]{V} \cdot n\sqrt{n}}{\sqrt[12]{cos(cos(E))}}$ & --- & --- & $A \cdot \frac{(\theta e)^3 M \sqrt[3]{Vn_p}}{n^{4/3} T \gamma_{a}^2}$ & ---\\
\hline
\textbf{RMSE} & 15.914 & 16.184 & 11.816 & \textbf{5.98} & 16.349 & RFR\\
\hline
\textbf{$R^{2}$} & 0.368 & 0.346 & 0.651 & \textbf{0.911} & -1.206 & RFR\\ 
\hline
\end{tabular}}
\caption{\label{tab:cv_results} 10-fold cross-validation performance of SR models, other ML models, and the slack model. Bold values correspond to the best ML/SR models}
\end{table}

\begin{figure}[H]
\centering
\includegraphics[width=.32\textwidth]{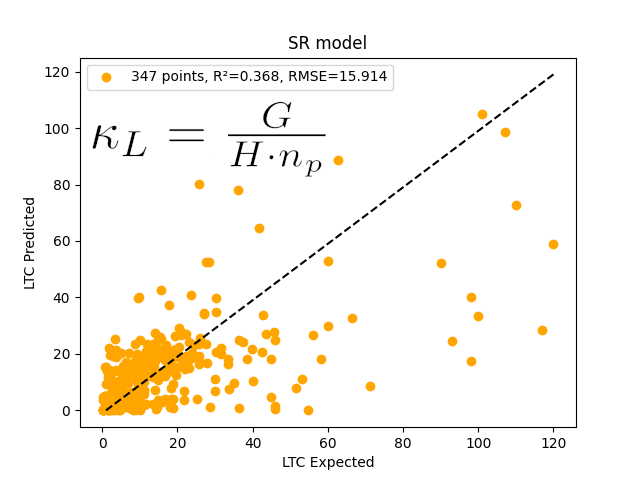}
\includegraphics[width=.32\textwidth]{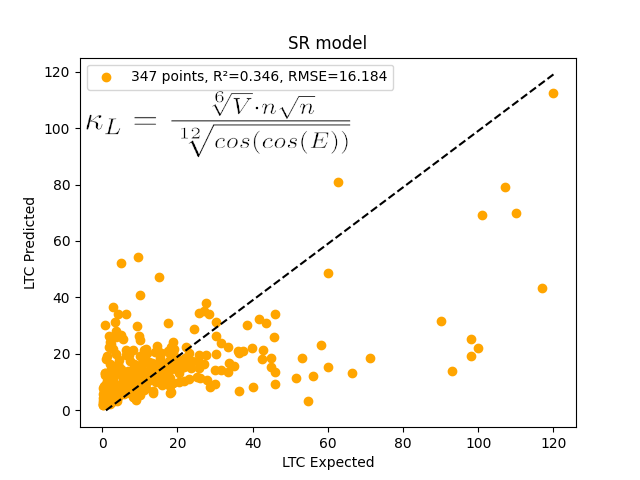}
\includegraphics[width=.32\textwidth]{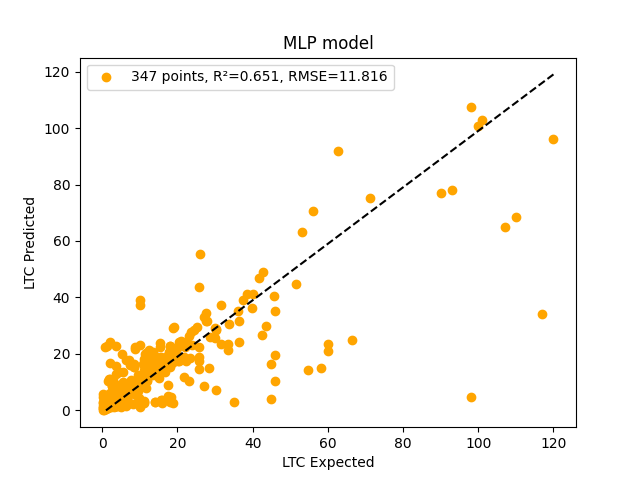}
\includegraphics[width=.32\textwidth]{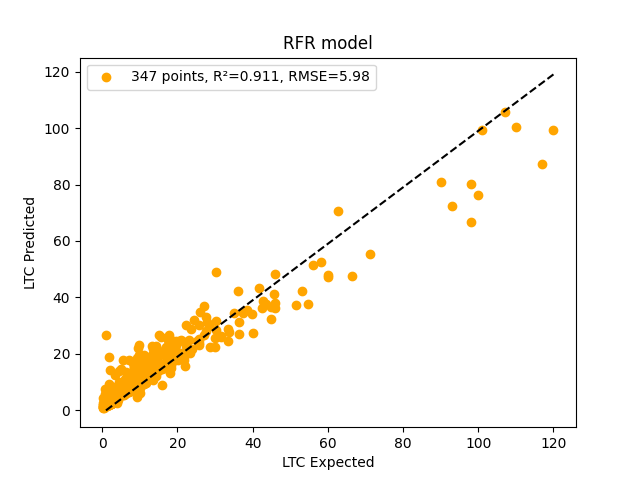}
\caption{Parity plots for the 10-fold cross-validation experiments of GP1, GP2, MLP, RFR}
\label{fig:cv_parity_plots}
\end{figure}

As shown in Figure \ref{fig:cv_parity_plots}, the SR models and the MLP and RFR models have all achieved good cross-validation prediction performance with $R^2$ scores of 0.368, 0.346, 0.651, and 0.911. In this evaluation approach, the samples are randomly shuffled and split into K=10 folds. Thus the test samples also have chance to find neighbor similar samples, thus good prediction performance. Overall, the Random Forest regressor has achieved the best performance with a $R^2$ of 0.911. Compared to its low extrapolation performance as shown in next section, the standard cross-validation is best for estimating interpolation performance.

\subsection{Extrapolation Testing Results}

Tables \ref{tab:extrapolation_results_top}, \ref{tab:extrapolation-topbottom}, \ref{tab:extrapolation_results_bottom}, and \ref{tab:extrapolation_results_middle} show the results from the four extrapolation testing sets.  The \emph{Best} column indicates the model that had the best performance on that set. All of the new formulae found in this section are analyzed in greater detail in Section \hyperref[sec:compare-formulae]{3.6}. The formulae displayed in the table have been simplified from the original forms created by the computational models, and thus may contain operators that exist outside of their associated function sets as defined in Table \ref{tab:sr_model_architectures}. These new operators are the result of combining operators used by the models. For example, $n_p \cdot n_p$ is simplified to $n_p^2$.



\begin{table}[H]
\centering
{\begin{tabular}{|c|c|c|c|c|c|c|} \toprule
  & \textbf{GP1} & \textbf{GP2} & \textbf{MLP} & \textbf{RFR} & \textbf{Slack} & \textbf{Best}\\
\hline
\textbf{Formula} & $\frac{Mn}{n_{p}}$ & $ln^{3}(|\sqrt[3]{Bcos(ln(|p|))}| \cdot |ln(|ln^{3}(|\sqrt{G}|)|)|)$ & --- & --- & $A \cdot \frac{(\theta e)^3 M \sqrt[3]{Vn_p}}{n^{4/3} T \gamma_{a}^2}$ & ---\\
\hline
\textbf{Training RMSE} & 21.458 & 19.759 & \textbf{18.792} & 20.680 & 20.697 & MLP\\
\hline
\textbf{Testing RMSE} & 5.089 & 5.558 & \textbf{4.166} & 6.327 & 4.727 & MLP\\ 
\hline
\textbf{Training $R^{2}$} & -0.059 & 0.102 & \textbf{0.188} & 0.016 & 0.053 & MLP\\ 
\hline
\textbf{Testing $R^{2}$} & \textbf{-38.381} & -45.97 & -45.97 & -59.871 & -32.971 & GP1\\
\hline
\end{tabular}}
\caption{\label{tab:extrapolation_results_bottom}Results after training the models on the top 80\% of the samples with the highest $\kappa_{L}$ values and testing on the bottom 20\% samples. Bold values correspond to the best ML/SR models}
\end{table}

\begin{table}[H]
\centering
{\begin{tabular}{|c|c|c|c|c|c|c|} \toprule
  & \textbf{GP1} & \textbf{GP2} & \textbf{MLP} & \textbf{RFR} & \textbf{Slack} & \textbf{Best}\\
\hline
\textbf{Formula} & $\frac{0.31B}{n_{p}}$ & $B^{(\frac{1}{2})}n_{p}^{\frac{1}{9}}cos(cos(sin(cos(V))))cos^{2}(\frac{1}{tan^3(e^{n})})$ & --- & --- & $A \cdot \frac{(\theta e)^3 M \sqrt[3]{Vn_p}}{n^{4/3} T \gamma_{a}^2}$ & ---\\
\hline
\textbf{Training RMSE} & 20.293 & 18.398 & \textbf{17.332} & 20.941 & 19.392 & MLP\\
\hline
\textbf{Testing RMSE} & \textbf{4.788} & 5.825 & 4.992 & 5.457 & 14.029 & GP1\\ 
\hline
\textbf{Training $R^{2}$} & 0.159 & 0.309 & \textbf{0.387 }& 0.105 & 0.036 & MLP\\ 
\hline
\textbf{Testing $R^{2}$} & \textbf{-6.718} & -10.424 & -7.391 & -9.027 & -65.266 & GP1\\
\hline
\end{tabular}}
\caption{\label{tab:extrapolation_results_middle}Results after training the models on the top and bottom 80\% of samples with lowest 40\% and highest 40\% $\kappa_{L}$ values and testing on the middle samples. Bold values correspond to the best ML/SR models}
\end{table}

\begin{table}[H]
\centering
{\begin{tabular}{|c|c|c|c|c|c|c|} \toprule
  & \textbf{GP1} & \textbf{GP2} & \textbf{MLP} & \textbf{RFR} & \textbf{Slack} & \textbf{Best}\\
\hline
\textbf{Formula} & $\frac{0.68Mn}{n_{p}}$ & $4ln^{2}(n) \cdot \sqrt[3]{Bn \cdot cos^{2}(\frac{15.49}{n})} \cdot \sqrt[9]{\frac{B}{sin^{2}(M)}}$ & --- & --- & $A \cdot \frac{(\theta e)^3 M \sqrt[3]{Vn_p}}{n^{4/3} T \gamma_{a}^2}$ & ---\\
\hline
\textbf{Training RMSE} & 5.534 & 4.729 & \textbf{3.113} & 4.070 & 19.451 & MLP\\
\hline
\textbf{Testing RMSE} & 45.257 & 44.038 & \textbf{43.979} & 45.640 & 37.999 & MLP\\ 
\hline
\textbf{Training $R^{2}$} & 0.196 & 0.413 & \textbf{0.745 }& 0.565 & 0.083 & MLP\\ 
\hline
\textbf{Testing $R^{2}$} & -1.98 & -1.821 & \textbf{-1.814} & -2.030 & -1.101 & MLP\\
\hline
\end{tabular}}
\caption{\label{tab:extrapolation_results_top}Results after training the models on the 80\% of the samples with the lowest $\kappa_{L}$ values and testing on the top 20\% samples. Bold values correspond to the best ML/SR models}
\end{table}

\begin{table}[H]
\centering
{\begin{tabular}{|c|c|c|c|c|c|c|} \toprule
  & \textbf{GP1} & \textbf{GP2} & \textbf{MLP} & \textbf{RFR} & \textbf{Slack} & \textbf{Best}\\
\hline
\textbf{Formula} & $\frac{G}{8.36B}$ & $\sqrt{Ecos(ln^{2}(|cos(ln(|cos(n_{p})|))|))}$ & --- & --- & $A \cdot \frac{(\theta e)^3 M \sqrt[3]{Vn_p}}{n^{4/3} T \gamma_{a}^2}$ & ---\\
\hline
\textbf{Training RMSE} & 8.396 & 7.206 & \textbf{5.423} & 6.623 & 18.808 & MLP\\
\hline
\textbf{Testing RMSE} & 42.99 & \textbf{40.869} & 40.922 & 42.839 & 36.101 & GP2\\ 
\hline
\textbf{Training $R^{2}$} & -0.01 & 0.256 & \textbf{0.579} & 0.372 & 0.124 & MLP\\ 
\hline
\textbf{Testing $R^{2}$} & -0.353 & \textbf{-0.223} & -0.226 & -0.334 & 0.046 & GP2\\
\hline
\end{tabular}}

\caption{\label{tab:extrapolation-topbottom}Results after training the models on the 80\% samples with middle $\kappa_{L}$ values and testing on the top and bottom 10\% of samples. Bold values correspond to the best ML/SR models}
\end{table}

\subsubsection{Performance comparison of SR and ML models to Slack model}
\label{sec:compare-sr-and-slack}

First, Table \ref{tab:extrapolation_results_bottom} shows the prediction performance of the algorithms when trained with top 80\% samples and tested on bottom 20\% samples. Over the training sets, the MLP model achieves the best performance with RMSE of 18.792 and $R^2$ of 0.188. On the testing set, the MLP model outperforms the GP1 model in RMSE (4.166 vs 5.089), though not in $R^{2}$ (-45.97 vs -38.381), indicating that the neural network's predictions have more variance than the function that GP1 produces.  It should be noted that the coefficient of determination $R^2$ can become negative when evaluated over test sets which are not included in the training set. GP2 is able to obtain 0.938 less RMSE than the Slack model on the training set, but 0.831 more error on the testing set, which shows that them model learned ungeneralizable trends in the subset of the data that it was shown. Strangely, while GP2 is evolved with more evaluations than the GP1 model and had access to more GP functions than GP1, the GP1 model achieved better metrics on the testing set than the GP2 model did. We attribute this incongruity to the larger search space that the GP2 model must navigate due to its larger pool of genetic programming functions.

We observe similar performance advantages of SR models compared to MLP, RFR, and Slack models in Table \ref{tab:extrapolation_results_middle}, which shows the performance of models when trained with top 40\% and bottom 40\% samples and tested on the middle 20\% samples. The MLP model is able to outperform all of the other models when evaluated against the set of data that it was trained upon; however, the GP1 model is superior on the testing set. This demonstrates that the GP1 model was able to learn trends from extremely poor and extremely successful thermal conductors, and accurately apply those trends to gain insight on the materials that lie in between those extremes. Interestingly, the GP1 model and RFR model are the only models that performed worse than the Slack model on the training set -- both GP2 and MLP were able to outperform the Slack model on the training set. The GP2 model is able to outperform the Slack Berman model across all metrics on all subsets of the dataset. It has a spectacularly better performance than the Slack model on the testing set, and performs better on the training set as well.

For the other two extrapolation experiments shown in Table \ref{tab:extrapolation_results_top} and Table \ref{tab:extrapolation-topbottom}, the results are a little bit different. Table \ref{tab:extrapolation_results_top} shows the results of models trained with bottom 80\% samples and tested on the top 20\% samples while Table \ref{tab:extrapolation-topbottom} shows the results of models trained with middle 80\% samples and tested on the two-ends 20\% samples. In Table \ref{tab:extrapolation_results_top}, the MLP model performs extraordinarily well. It is able to outperform all other SR/ML models across all of the testing metrics, and obtains 6.25x lower RMSE than the Slack model on the training set. Unfortunately, it underperforms on the testing set, demonstrating less aptitude than the Slack model for predicting materials at the upper end of the $\kappa_{L}$ spectrum. In Table \ref{tab:extrapolation_results_top}, the best SR model GP2 achieves a RMSE of 44.038 and $R^2$ of -1.821 over the test set which is worse than the RMSE of 37.999 and $R^2$ of -1.101 of the Slack model. In Table \ref{tab:extrapolation-topbottom}, the GP2 model outperforms all of the other ML/SR models on the testing set, though it still underperforms when compared to the Slack model. Further attention to Table \ref{tab:extrapolation_results_top} and Table \ref{tab:extrapolation-topbottom} reveals that when trained on the middle 80\%, the GP1 model produces a formula with an 139.48\% increase in $R^2$ score as compared to its score in Table \ref{tab:extrapolation_results_top}. Similarly, GP2 sees a 156.36\% increase in its own $R^2$ score. As expected, this demonstrates that training the model on a diverse set of data points yields increased extrapolative power. Naturally, it should follow that the most diverse training set should provide the most extrapolative potential. Table \ref{tab:extrapolation_results_middle} shows that this is true.

As noted above, training models on a diverse set of data provides the most extrapolative potential as opposed to alternative methods. Naturally, it follows that training the model on the bottom 40\% of the dataset and top 40\% of the dataset would yield the most accurate formulae, as the SR models would be exposed to examples of both materials with high and low $\kappa_{L}$. This is supported by Table \ref{tab:extrapolation_results_middle}, in which both GP2 and GP1 yield formulae that outperform the Slack-Berman equation. On the training set, GP1 and GP2 perform comparably to the Slack model, with GP1 estimating $\kappa_{L}$ with 0.901 more RMSE and GP2 estimating $\kappa_{L}$ with 0.994 less RMSE. However, on the testing sets, GP1 and GP2 demonstrate that they are significantly more accurate. GP1 has 9.241 RMSE less than the Slack model on the testing set, and GP2 achieves an RMSE of 5.825, 8.204 less than the Slack model on the same set (14.029). The two formulae produced by GP1 and GP2 perform similar to the Slack model on the training set, but are able to predict the median 20\% of materials with 2.93x and 2.4x less error than the Slack model, despite the fact that they have never seen materials with $\kappa_{L}$ in that range before. This successful prediction proves that the SR models are not overfitting to latent trends in their training sets, but have uncovered relationships that govern the calculation of $\kappa_{L}$. Formula \ref{eq:gp1-middle-formula} represents the formula generated by GP1 from this training set, and Formula \ref{eq:gp2-middle-formula} represents that generated by GP2. We discuss these formulae further in Section \hyperref[sec:compare-formulae]{3.6}, but for convenience we provide them here.

Now there is one remaining question: Why do the SR models work better when trained with the top 80\% and tested on the bottom 20\% compared to when they are trained with the bottom 80\% and tested on the top 20\%? After close inspection of the sample distribution in Figure \ref{fig:distribution}, it seems that this is caused by the extremely sparse amount of samples in the high-$\kappa_{L}$ area compared to the dense amount of samples in the bottom $\kappa_{L}$ area. As a result, whenever the test set includes the top $\kappa_{L}$ area, the extrapolation performance will be very low. This result confirms the importance of training ML and SR models with balanced diverse data samples.



\begin{equation}
    \kappa_{L} = \frac{0.31B}{n_{p}}
    \label{eq:gp1-middle-formula}
\end{equation}

\begin{equation}
    \kappa_{L} = B^{(\frac{1}{2})}n_{p}^{\frac{1}{9}}cos(cos(sin(cos(V))))cos^{2}(\frac{1}{tan^3(e^{n})}) \label{eq:gp2-middle-formula}
\end{equation}

\subsubsection{Performance comparison of SR, RFR and MLP}
\label{sec:compare-sr-and-others}

Comparing the SR and machine learning models performance against each other on the datasets reveals a few interesting results. While the MLP model is very effective at learning from the data it is shown, it does not have the same extrapolative potential that the SR models have. MLP outperforms all other models on the training sets when evaluated by both RMSE and $R^2$, as shown by Tables \ref{tab:extrapolation_results_top}, \ref{tab:extrapolation-topbottom}, \ref{tab:extrapolation_results_bottom} and \ref{tab:extrapolation_results_middle}. However, it is not always able to outperform the GP models on the testing sets. The MLP model's predictions on the validation sets are close to that of the SR models, with the largest difference in RMSE being 2.068 in Table \ref{tab:extrapolation-topbottom}.  The fact that it is unable to consistently match the Symbolic Regression models' performance on the testing sets yet outperforms them on the training sets demonstrates that while the neural network is excellent at creating a model that accurately predicts materials in the range it has seen before, it does not always transfer this knowledge to materials outside of this range.

The RFR model does not perform better than the MLP model for any metric on any set of materials from the dataset. However, it does obtain lower RMSE and higher $R^2$ values than the SR models on the training sets for Tables \ref{tab:extrapolation_results_top} and \ref{tab:extrapolation-topbottom}. Despite this, it is unable to compare with the SR models on any of the validation sets except for Table \ref{tab:extrapolation_results_middle} and Table \ref{tab:extrapolation-topbottom}, where it outperforms one of the SR models but is surpassed by the other SR model. For example, in Table \ref{tab:extrapolation_results_middle}, the RFR model outperforms the GP2 formula's RMSE by 0.368, but is worse than GP1's formula by 0.669. None of the models were able to perform better than the Slack model on the extrapolation sets depicted in Table \ref{tab:extrapolation_results_top} and Table \ref{tab:extrapolation-topbottom}. However, this is not to say that the Slack formula is superior for calculating the $\kappa_{L}$ of materials in those validation sets; it simply means that the models needed data points from those sets in order to learn the relationships for them. The models demonstrated their efficacy for $\kappa_{L}$ approximation on those materials through the RMSE reported in Tables \ref{tab:extrapolation_results_bottom} and \ref{tab:extrapolation_results_middle}, where they demonstrated comparable performance to the Slack formula.

\subsubsection{Performance comparison of GP1 and GP2}
\label{sec:compare-sr-models}

As shown in Table \ref{tab:sr_model_architectures}, we evaluated two symbolic regression algorithms to evolve SR models: GP1 and GP2, where GP1 corresponds to a simpler function set with max tree height of 7, leading to simpler models. On the other hand, GP2 model is trained with more complex function set with a tree height of 10, leading to more complex models. 

All the GP1 and GP2 performance results with four extrapolation experiments are shown in Table \ref{tab:extrapolation_results_top}, Table \ref{tab:extrapolation-topbottom}, Table \ref{tab:extrapolation_results_middle}, and Table \ref{tab:extrapolation_results_bottom}. We find that the GP1 model outperforms the GP2 model when the top portion of the dataset is included in the testing set (Table \ref{tab:extrapolation_results_bottom} and Table \ref{tab:extrapolation_results_middle}). This can be noted by observing Table \ref{tab:extrapolation_results_bottom}, in which GP1's RMSE (5.089) is 0.469 less than GP2's RMSE of 5.558. However, when the top section of the dataset is excluded from the training set (as in Table \ref{tab:extrapolation_results_top} and Table \ref{tab:extrapolation-topbottom}), the GP2 model outperforms the GP1 model. This can be observed by comparing GP1's RMSE score of 42.99 to GP2's RMSE of 40.869 in Table \ref{tab:extrapolation-topbottom}. 


\subsection{Parity Plot Analysis}
\begin{figure}[H]
\centering
\includegraphics[width=.45\textwidth]{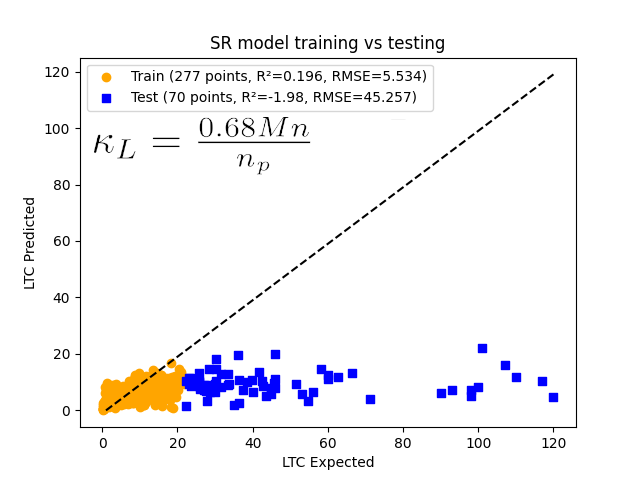}
\includegraphics[width=.45\textwidth]{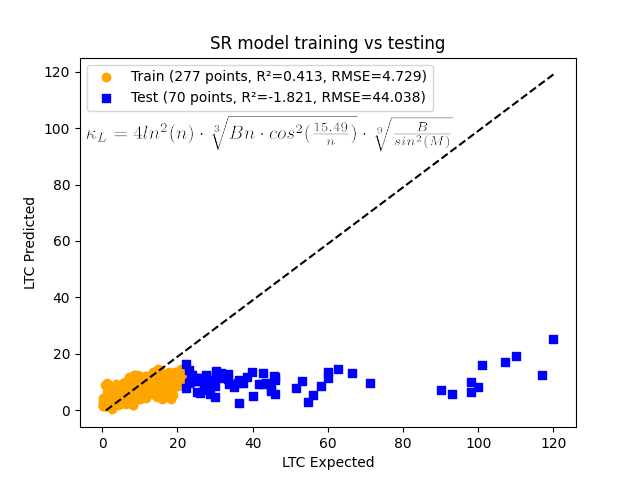}
\includegraphics[width=.45\textwidth]{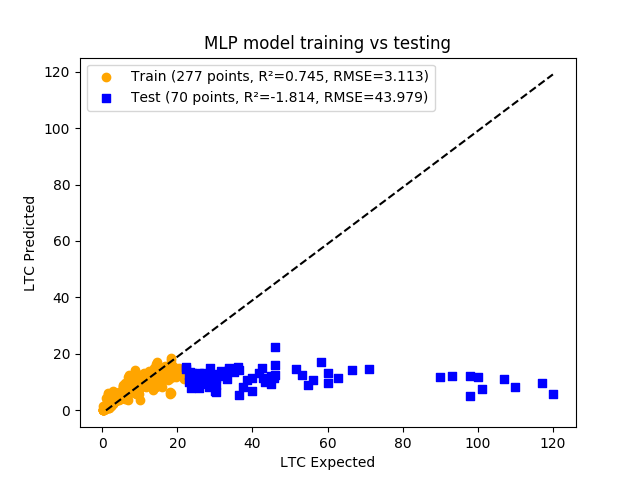}
\includegraphics[width=.45\textwidth]{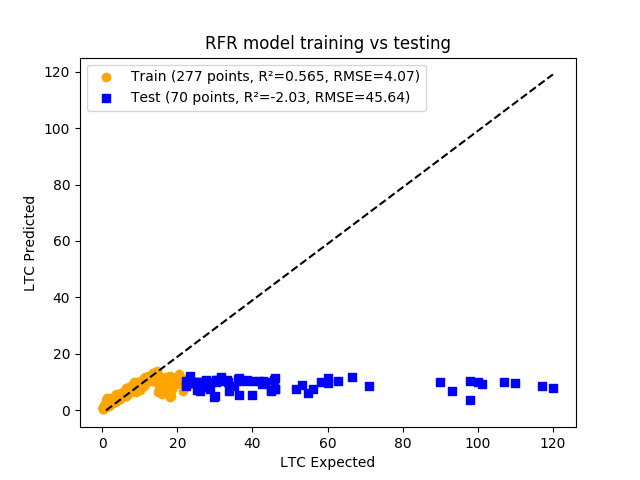}
\includegraphics[width=.45\textwidth]{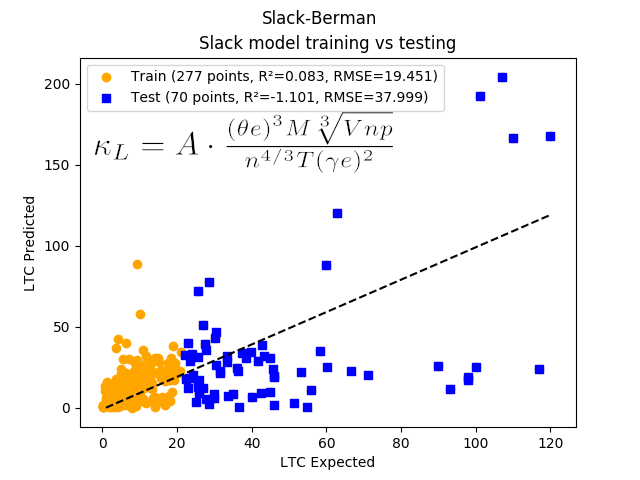}
\caption{Parity plots for the GP1, GP2, MLP, RFR, and Slack-Berman models on the top 20\% extrapolation testing set.}
\label{fig:top_parity_plots}
\end{figure}

\begin{figure}[H]
\centering
\includegraphics[width=.45\textwidth]{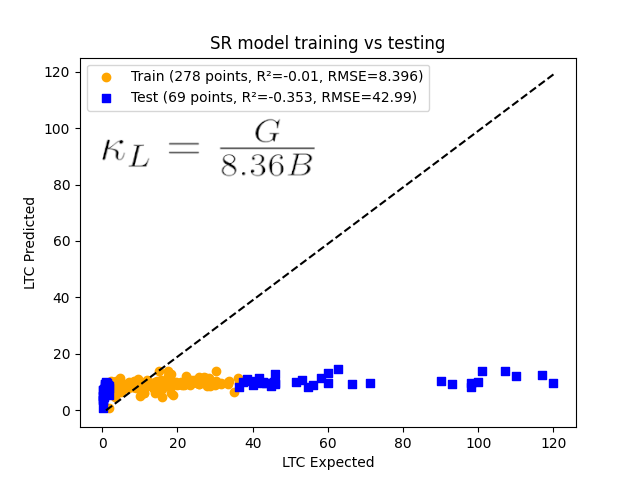}
\includegraphics[width=.45\textwidth]{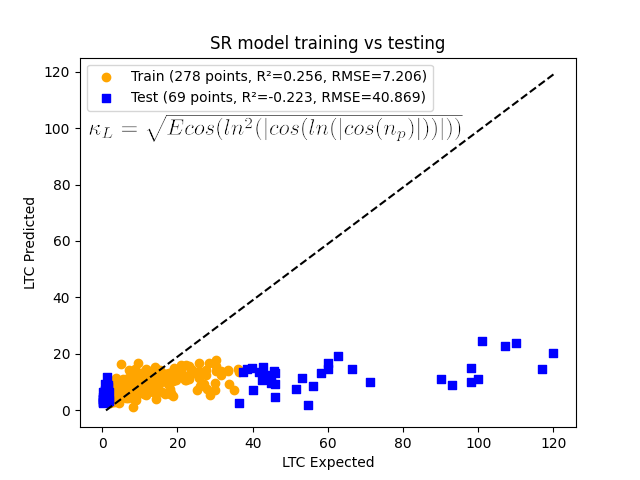}
\includegraphics[width=.45\textwidth]{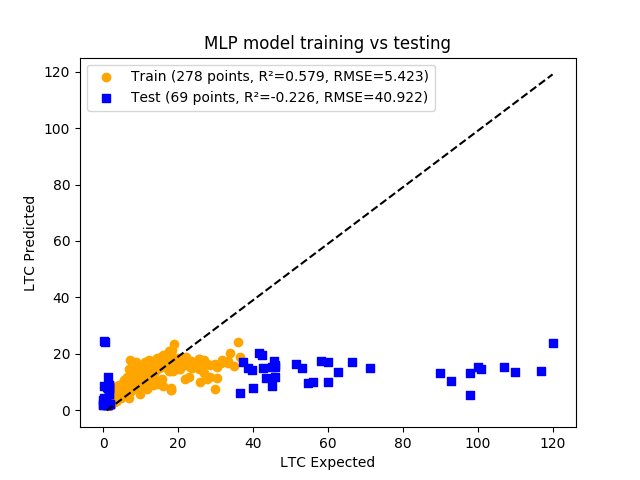}
\includegraphics[width=.45\textwidth]{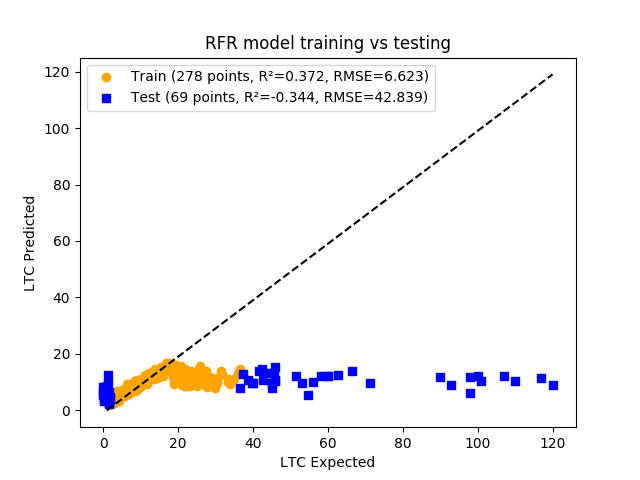}
\includegraphics[width=.45\textwidth]{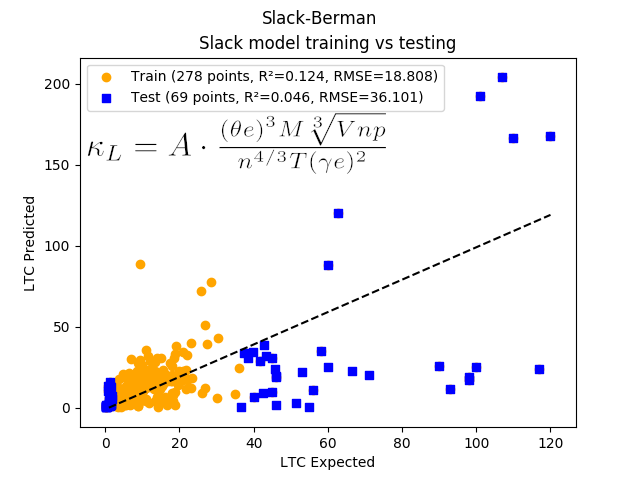}
\caption{Parity plots for the GP1, GP2, MLP, RFR, and Slack-Berman models on the top 10\% and bottom 10\% extrapolation testing set.}
\label{fig:top_bottom_parity_plots}
\end{figure}

\begin{figure}[H]
\centering
\includegraphics[width=.45\textwidth]{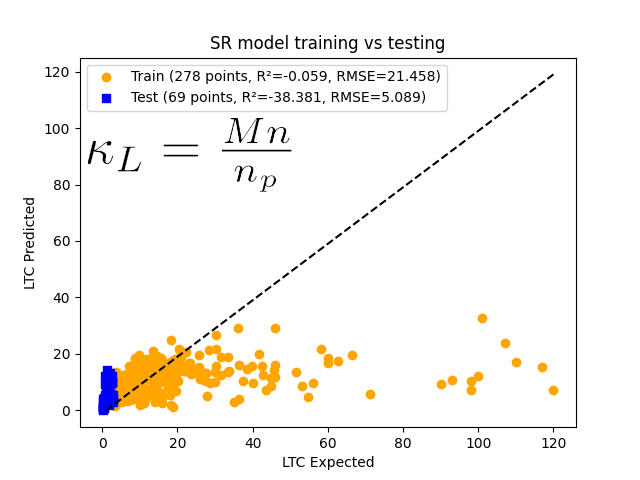}
\includegraphics[width=.45\textwidth]{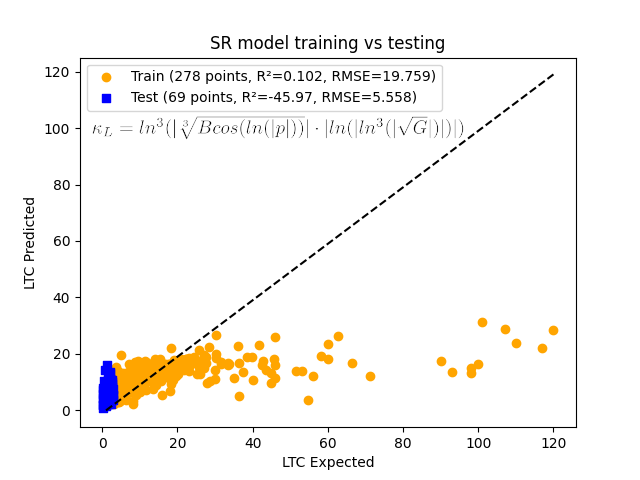}
\includegraphics[width=.45\textwidth]{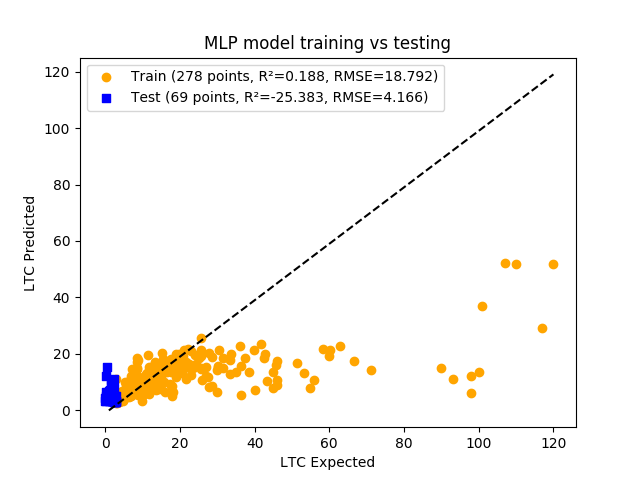}
\includegraphics[width=.45\textwidth]{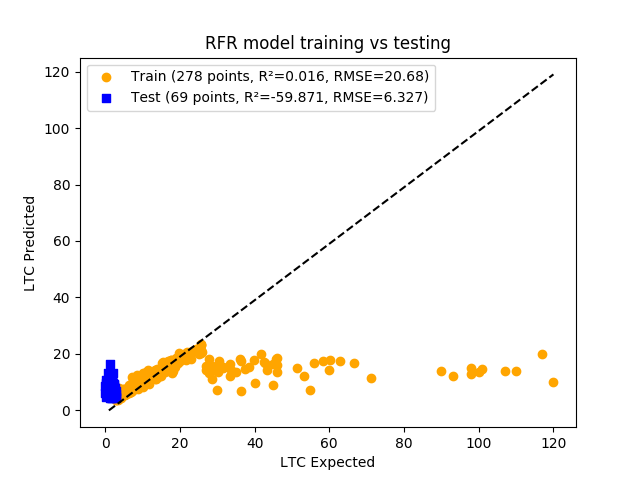}
\includegraphics[width=.45\textwidth]{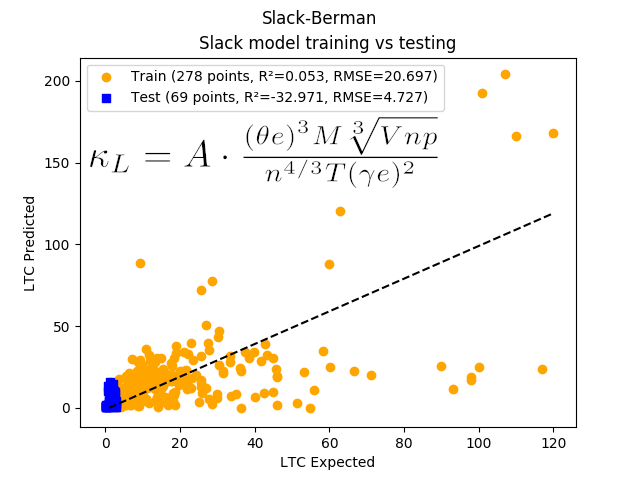}
\caption{Parity plots for the GP1, GP2, MLP, RFR, and Slack-Berman models on the bottom 20\% extrapolation testing set.}
\label{fig:bottom_parity_plots}
\end{figure}

\begin{figure}[H]
\centering
\includegraphics[width=.45\textwidth]{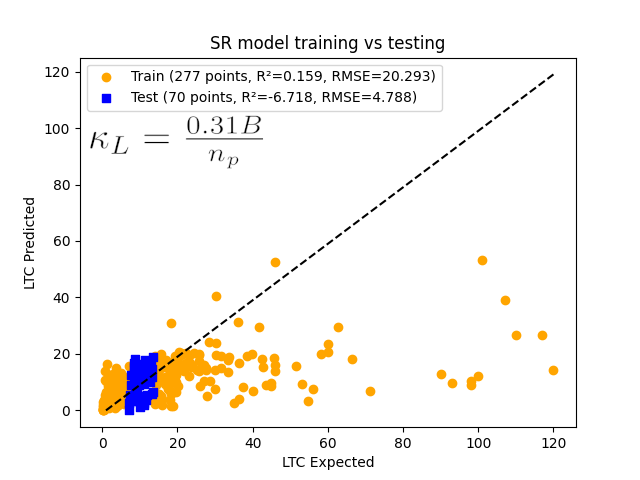}
\includegraphics[width=.45\textwidth]{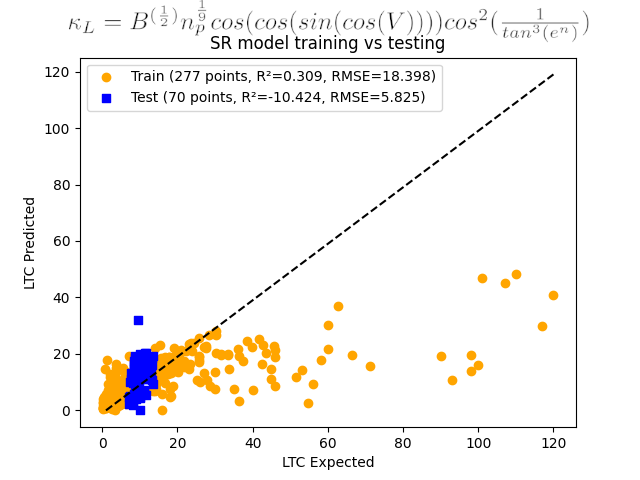}
\includegraphics[width=.45\textwidth]{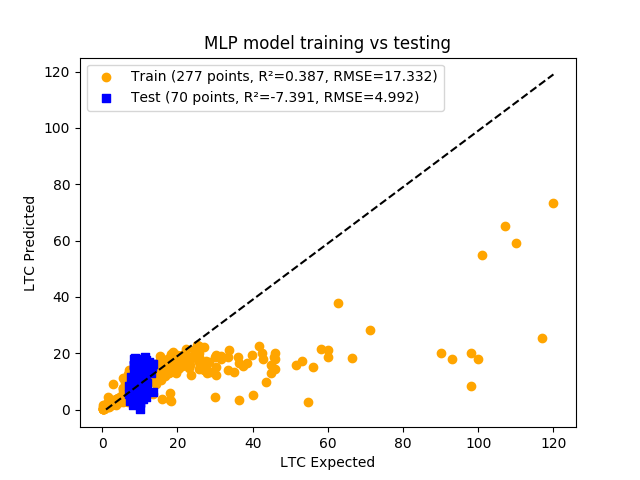}
\includegraphics[width=.45\textwidth]{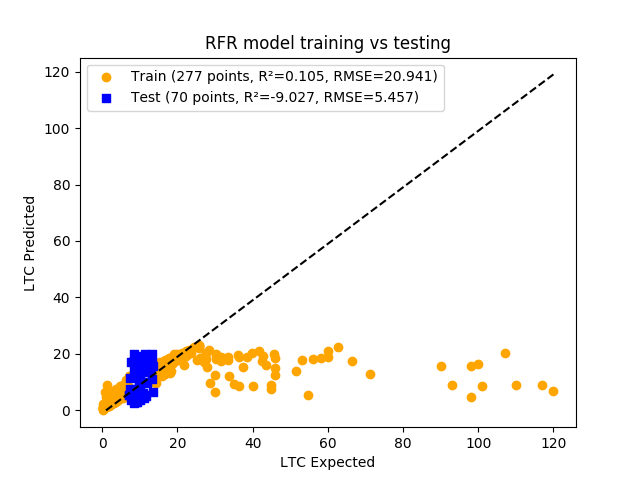}
\includegraphics[width=.45\textwidth]{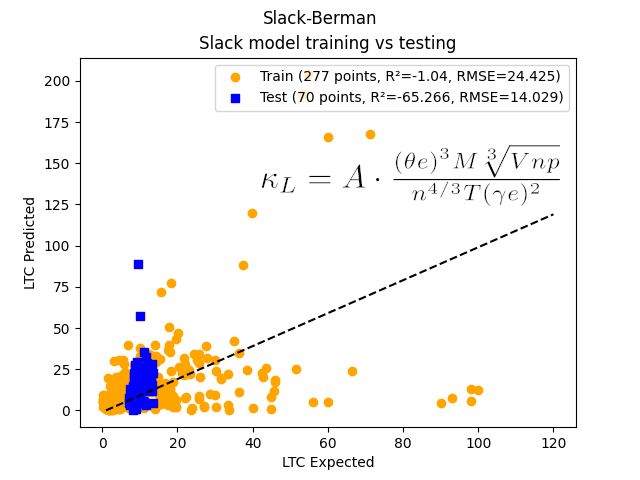}
\caption{Parity plots for the GP1, GP2, MLP, RFR, and Slack-Berman models on the middle 20\% extrapolation testing set.}
\label{fig:middle_parity_plots}
\end{figure}

To further understand why the SR and ML models have unexpected low extrapolation prediction performance, we create a set of parity plots (Figures \ref{fig:top_parity_plots} - \ref{fig:middle_parity_plots}) for all the four extrapolation experiments and aim to figure how the sample distribution affects the prediction performance of ML and SR models. In all of the plots, the orange points represent training samples while the blue points are test samples. However, for the Slack model, both colors represent testing points. This is because the Slack model is an empirical model, and thus does not require training.

Firstly, we find that compared to the random cross-validation performance results(Figure~\ref{fig:cv_parity_plots}, the extrapolation prediction performances of all SR and ML models are unexpectedly low, consistent with our previous observations \cite{xiong2020evaluating} along with other analysis\cite{fannjiang2020autofocused} on out-of-distribution generalization issues. 

Second, across all the parity plots, there is a clear propensity for the prediction models to underestimate $\kappa_{L}$ (most of the samples are below the diagonal line). This is not an unexpected development, as the dataset contains many more materials on the lower spectrum of $\kappa_{L}$ materials than the upper bound. As Figure \ref{fig:distribution} demonstrates, its distribution is positively skewed. The models that were trained on materials with lower $\kappa_{L}$ values often underestimate the values of their testing sets, as shown in Figures \ref{fig:top_parity_plots}. The models trained on the upper side of the material $\kappa_{L}$ spectrum tend to generate overestimates, as demonstrated by Figure \ref{fig:bottom_parity_plots}. In Figure \ref{fig:top_bottom_parity_plots}, the models both over-approximate the lower $\kappa_{L}$ materials and under-estimate the higher $\kappa_{L}$ materials' values, as the models shown in that figure were trained on the middle 80\% of the dataset. Finally, Figure \ref{fig:middle_parity_plots} shows a more even balance, with variance both above and below the $y=x$ line. This is due to the fact that the models were provided a diverse training set of both high and low $\kappa_{L}$ materials, as previously noted in Section \hyperref[sec:compare-sr-and-slack]{3.3}. They still show a tendency to underpredict the values on the upper bound, with the SR models demonstrating a slightly more standard yet still skewed variance.

The parity plots (Figures \ref{fig:top_parity_plots} - \ref{fig:middle_parity_plots}) for all extrapolation sets indicate that the models behaved largely as expected on their training sets; on some materials, the models overestimated the $\kappa_{L}$, and on others, they underestimated the $\kappa_{L}$. The RFR model's parity plots have an interesting spread on the training sets. Most easily seen in Figure \ref{fig:bottom_parity_plots}, the RFR model predicts the lower end of its training sets with relatively low error, but then abruptly begins predicting nearly the same $\kappa_{L}$ for all of its materials with some variation. This variation is lowest in Figure \ref{fig:top_parity_plots}, and highest in Figure \ref{fig:middle_parity_plots}. The point at which the RFR model experiences its estimation accuracy falloff occurs at a logical point in each of its parity plots. In Figure \ref{fig:top_parity_plots}, the jump in error occurs relatively early in the plot, whereas in Figures \ref{fig:bottom_parity_plots} and \ref{fig:middle_parity_plots} it occurs later. This is because Figure \ref{fig:top_parity_plots} contains materials of low $\kappa_{L}$, so the error spikes when the $\kappa_{L}$ rises. This same spike occurs later in the other figures, because they include member nodes of higher $\kappa_{L}$ values for the model to learn from. The RFR models' parity plot testing sets continue trends identified by the models for the higher values in their training sets, which indicates that the models have found similarities in the fields of the two subsets. This sudden spike in performance is not an unexpected development, as the RFR model is decision tree based  \cite{xiong2020evaluating}.

\subsection{Formula Comparison}
\label{sec:compare-formulae}

Comparing the formulae generated by the SR models yields some interesting insight, particularly when they are compared to the Slack-Berman formula (\ref{eq:slack-berman}). Both formulae produced by the SR model in the extrapolation experiments (Equations \ref{eq:gp1-middle-formula} and \ref{eq:gp2-middle-formula}) place the $B$ field in the numerator, indicating that $\kappa_{L}$ scales with bulk modulus. This is consistent with current kinetic theory of phonon transport that the inclusion of bulk modulus as a variable in the formula is essential to approximating a material's $\kappa_{L}$.

Very interestingly, Equations \ref{eq:gp1-kfold-formula}
 and \ref{eq:gp2-kfold-formula} do not make use of the $B$ field whatsoever. This contradicts the aforementioned kinetic theory of phonon transport. In addition, the only variable that Equation \ref{eq:gp1-kfold-formula} has in common with Equations \ref{eq:gp1-middle-formula} and \ref{eq:gp2-middle-formula} is $n_{p}$. Equation \ref{eq:gp2-middle-formula} and Equation \ref{eq:gp2-kfold-formula} have some overlap in that they both make use of the $V$ and $n$ variables.
 
Equation \ref{eq:gp2-middle-formula} places the $n_p$ variable in the numerator, which corresponds to the Slack formula's usage of the field. However, Equation \ref{eq:gp1-middle-formula} places it in the denominator. This disparity indicates a disagreement between the formulae, where Equation \ref{eq:gp2-middle-formula} assumes that $\kappa_{L}$ has a positive correlation with the number of atoms in a primitive cell, and Equation \ref{eq:gp1-middle-formula} indicates that they have a negative correlation. We conclude that Equation \ref{eq:gp2-middle-formula}'s usage of the field is most likely correct, as it boasts a lower RMSE than Equation \ref{eq:gp1-middle-formula} (16.643 vs 18.255).

Perhaps \hl{the} most interesting of all is the set of variables selected by the formulae. The Slack formula makes use of 6 variables and 2 constants ($A \& T$), whereas the most accurate formula that our models produced (Formula \ref{eq:gp2-kfold-formula}) uses only three variables and achieves a higher accuracy. The two formulae have the $n$ and $V$ variables in common.

While the models produced a multitude of potential formulae, we have elected to only include those with the least error in the primary section of this work. A selection of other noteworthy formulae have been collected based on their interesting properties, and have been included in the Appendix.

\begin{equation}
    \kappa_{L} = \frac{0.31B}{n_{p}} \tag{\ref{eq:gp1-middle-formula}}
\end{equation}

\begin{equation}
    \kappa_{L} = B^{(\frac{1}{2})}n_{p}^{\frac{1}{9}}cos(cos(sin(cos(V))))cos^{2}(\frac{1}{tan^3(e^{n})}) \tag{\ref{eq:gp2-middle-formula}}
\end{equation}

\begin{equation}
    \kappa_{L} = \frac{G}{H \cdot n_{p}}
    \label{eq:gp1-kfold-formula}
\end{equation}

\begin{equation}
    \kappa_{L} = \frac{\sqrt[6]{V} \cdot n\sqrt{n}}{\sqrt[12]{cos(cos(E))}}
    \label{eq:gp2-kfold-formula}
\end{equation}

\subsection{Discussion}

All of the machine learning models reviewed in this study have their own unique advantages and disadvantages to their use. Symbolic Regression is computationally expensive and time consuming during the training stage, but it leads to formulae that are physically meaningful and have enhanced extrapolative capacity and also run fast during the prediction stage. Random Forest reduces overfitting and variance through the usage of bagging and ensemble learning. Multilayer Perceptron neural networks are able to accurately discover nonlinear relationships from training data, and with a large enough dataset, they are able to use this information to estimate data points that lie outside their training set.

Using RMSE and $R^2$ as metrics for evaluation, the Symbolic Regression models used in this work were collectively more effective than any other model on the extrapolation validation sets. The MLP model performed comparably on a number of validation sets, and outperformed the GP1 and GP2 models on some others, but overall it was less effective on the validation sets. In addition, the SR models provide formulae that can be analyzed to obtain physical insight into the relationships of the variables in the formulae; Multilayer Perceptron and Random Forest models cannot provide the same level of insight.

Our Symbolic Regression models produced formulae that are able to calculate $\kappa_{L}$ with comparable or greater accuracy than the traditional Slack formula (Equation \ref{eq:slack-berman}), all while using less variables to do so. We demonstrate the validity of our Symbolic Regression methodology by showing that it can approximate the Slack formula with an $R^2$ score of 0.946 (Fig \ref{fig:sr_reproduction_parity_plot}). There are a multitude of other sources that have proven the Symbolic Regression algorithm's capacity for discovering physical laws \cite{Hernandez2019}\cite{Schmidt81}. Symbolic Regression provides computers with the ability to discover natural laws from raw data, and even bring insights to boot. Formulae \ref{eq:gp1-middle-formula} and \ref{eq:gp2-middle-formula} successfully reproduces the physical insight, although already known, that $\kappa_{L}$ scales positively with bulk modulus. Formulae \ref{eq:gp1-middle-formula} and \ref{eq:gp1-kfold-formula} reproduce the physical insight that $\kappa_{L}$ scales negatively with the number of atoms in the primitive unit cell of a material.

In addition to the other models discussed in the paper, we also ran the lastest Symbolic Regression algorithm, the AI-Feynman algorithm \cite{udrescu2020ai} over our dataset using the implementation in the github repository by Udrescu \cite{udrescu} \cite{udrescu2020ai20}. Initially, the algorithm did not converge to any useable formula because our dataset contained too many input variables. However, even after we restricted the dimensionality of the problem to only the six variables used by the orginal Slack model and allowed the model to run continuously for nine days, it still did not converge. The AI-Feynman algorithm on paper is a very strong candidate for predicting formulae for LTC, as it is a symbolic regression algorithm that does not rely on genetic programming. Rather than using an evolutionary algorithm, the AI-Feynman algorithm uses neural networks to simplify the data it is provided with before using a brute force algorithm to try all symbolic expressions possible in order of ascending complexity. The algorithm is very promising and has the capability to exploit the units of variables. Unfortunately, with our currently limited dataset we were unable to successfully apply it to get better formulae.

There are a few areas in which this study could be improved. Firstly, collating a larger dataset of materials with measured $\kappa_{L}$ values will enable all three types of models explored in this study to obtain lower error metrics and resolve issues with variance and bias from the models. Obtaining a balanced dataset that has a normally distributed range of $\kappa_{L}$-ranked materials will permit the models to improve their performance across all categories, especially when predicting materials with high $\kappa_{L}$. Outside of changes to the dataset, the symbolic regression methodology has some room for improvement. Attributing units and types to the variables in the dataset before feeding them to the SR models will allow for the inclusion of binary operators that require consistent units, such as addition and subtraction. The inclusion of these operators would substantially increase the hypothesis space of the SR models, potentially leading to more accurate models.

Even with these limitations, Symbolic Regression has demonstrated that it can learn from raw experimental data and intelligently produce equations and formulae that can predict unseen values. In this work, we have proven that genetic programming has the capacity to create formulae that are more accurate and more consistent than models that have been derived by physicists for the same task (Formula \ref{eq:slack-berman}). It can infer relationships that are relevant to materials outside of the range that it was trained on, and it does so with less error than neural networks and random forest regressors trained on the exact same data.


\vspace{6pt} 



\section{contributon}

 Conceptualization, J.H. and M.H.; methodology, J. H. and C.L.; software, C.L., J.H. and Y.Z.; validation, C.L., J.H. and M.H.; investigation, C.L., J.H. and M.H. ; resources, J.H.; data curation, K.Y. and M.H.; writing--original draft preparation, C.L and J.H.; writing--review and editing, C.L., J.H., Y.Z., K.Y. and M.H.; visualization, C.L.; supervision, J.H.; project administration, J.H.; funding acquisition, J.H. and M.H.



\section{Acknowledgments}

This research was supported, in part, by a grant from the Magellan Scholar program, from the Department of Undergraduate Research at the University of South Carolina, Columbia. Research reported in this publication was also partially supported by the National Science Foundation under grant numbers: 1940099, 1905775, OIA-1655740 (via SC EPSCoR/IDeA 20-SA05) and by DOE under grant number DE-SC0020272.

\section{Confict of Interest Declarations} 
The authors declare no conflict of interest.
\appendix
\textbf{Appendix}

\section{Noteworthy Formulae from Symbolic Regression Models}

The Symbolic Regression models generated a large quantity of formulae during the training phase. Ultimately, most were uninteresting. While we elected to only include the best performing formulae in the main body of this work, it would be neglectful to omit several of the more physically interesting formulae that were found by the models. Below are some of the aforementioned interesting candidates, and their average MAE and $R^{2}$ scores across both the training and validation sets.

\begin{equation}
    \kappa_{L} = 3.55 \cdot \sqrt{\frac{B}{(n_p)^2 \cdot \sqrt{H}}} \label{eq:a1}
\end{equation}

\begin{equation}
    \kappa_{L} = \frac{B}{H \cdot n_p^{2}} \label{eq:a2}
\end{equation}

\begin{equation}
    \kappa_{L} = \frac{E \cdot v}{n_p} \label{eq:a3}
\end{equation}

\begin{equation}
    \kappa_{L} = \frac{B}{n_p^{2}} \label{eq:a4}
\end{equation}

\begin{equation}
    \kappa_{L} = \frac{G^{\frac{2}{3}} ln(2.60)}{ln(|n_p|) H^{\frac{1}{9}}} \label{eq:a5}
\end{equation}

\begin{equation}
    \kappa_{L} = \sqrt[6]{\frac{B^{3} \cdot \sqrt[H]{e} \cdot \sqrt{ln(0.45 \cdot |n|)}}{H}} \label{eq:a6}
\end{equation}

\begin{table}[H]
    \centering
    \begin{tabular}{|c|c|c|} \toprule
         Formula & MAE & $R^2$ \\
         \hline
         Slack-Berman & 10.62 & 0.078 \\
         \hline
         \ref{eq:a1} & 9.461 & 0.135 \\
         \hline
         \ref{eq:a2} & 9.373 & 0.215 \\
         \hline
         \ref{eq:a3} & 9.360 & 0.175 \\
         \hline
         \ref{eq:a4} & 8.916 & 0.234 \\
         \hline
         \ref{eq:a5} & 9.303 & 0.210 \\
         \hline
         \ref{eq:a6} & 9.927 & 0.091 \\
         \hline
    \end{tabular}
    \caption{Noteworthy formulae MAE \& $R^2$}
    \label{tab:my_label}
\end{table}

{\small
\bibliographystyle{corlabbrvnat}
\bibliography{references.bib}
}

\end{document}